\documentclass[10pt,twocolumn,letterpaper]{article}

\usepackage[pagenumbers]{iccv} 

\definecolor{iccvblue}{rgb}{0.21,0.49,0.74}
\usepackage[pagebackref,breaklinks,colorlinks,allcolors=iccvblue]{hyperref}

\def\paperID{5654} 
\def\confName{ICCV}
\def\confYear{2025}

\title{RGB-Phase Speckle: Cross-Scene Stereo 3D Reconstruction via Wrapped Pre-Normalization}

\author{
Kai Yang$^{1}$ \quad Zijian Bai$^{2}$ \quad Yang Xiao$^{1}$ \quad Xinyu Li$^{2}$ \quad Xiaohan Shi\\
$^{1}$ Southwest Jiaotong University, Chengdu, China\\
$^{2}$ Sclead, Chengdu, China
}

\begin{document}
\maketitle
\begin{abstract}
    3D reconstruction garners increasing attention alongside the advancement of high-level image applications, where dense stereo matching (DSM) serves as a pivotal technique. Previous studies often rely on publicly available datasets for training, focusing on modifying network architectures or incorporating specialized modules to extract domain-invariant features and thus improve model robustness. In contrast, inspired by single-frame structured-light phase-shifting encoding, this study introduces RGB-Speckle, a cross-scene 3D reconstruction framework based on an active stereo camera system, designed to enhance robustness. Specifically, we propose a novel phase pre-normalization encoding-decoding method: first, we randomly perturb phase-shift maps and embed them into the three RGB channels to generate color speckle patterns; subsequently, the camera captures phase-encoded images modulated by objects as input to a stereo matching network. This technique effectively mitigates external interference and ensures consistent input data for RGB-Speckle, thereby bolstering cross-domain 3D reconstruction stability. To validate the proposed method, we conduct complex experiments: (1) construct a color speckle dataset for complex scenarios based on the proposed encoding scheme; (2) evaluate the impact of the phase pre-normalization encoding-decoding technique on 3D reconstruction accuracy; and (3) further investigate its robustness across diverse conditions. Experimental results demonstrate that the proposed RGB-Speckle model offers significant advantages in cross-domain and cross-scene 3D reconstruction tasks, enhancing model generalization and reinforcing robustness in challenging environments, thus providing a novel solution for robust 3D reconstruction research.
\end{abstract}
\section{Introduction}
\label{sec:intro}

3D reconstruction involves extracting geometric information from images of the measured object. This technique finds extensive application across diverse fields, including industrial inspection, biomedical imaging, Simultaneous Localization and Mapping (SLAM) navigation, and optical image measurement\cite{LU2022106873,ZHU2023107542,VANCROMBRUGGE2020106305,VANDERJEUGHT201618}, owing to its non-contact sensing, rapidity, and high precision. The effectiveness of these advantages largely depends on the speed and robustness of the reconstruction algorithm employed. In this regard, 3D perception schemes primarily based on multi-frame structured light and binocular vision routes have emerged as popular approaches, inspired further advancements. 

\begin{figure}
    \centering
    \includegraphics[width=1\linewidth]{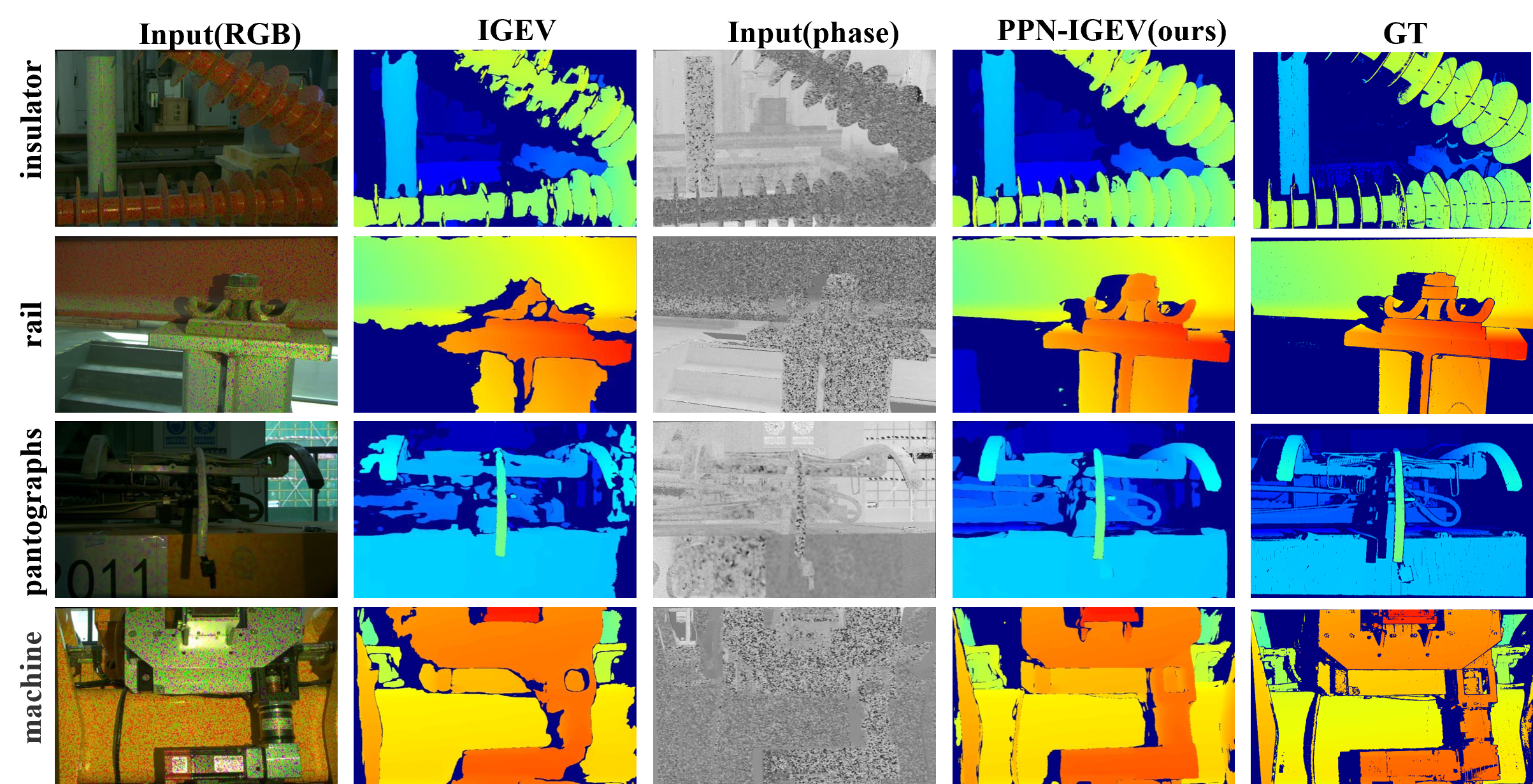}
    \caption{Comparison of the performance in challenging scenes. Columns from left to right denote sample input RGB image,the predicted disparities of the with rgb-speckle datasets,input phase image, the predicted disparities of the IGEV trained with phase pre-normalized images on the same datasets.}
    \label{fig:intro}
\end{figure}

However, the multi-frame projection method is restricted in scenarios involving motion or limited image capture time due to interference from vibrations and motion between image capture intervals. Single-frame methods represented by Fourier Transform Profilometry (FTP)\cite{Liu2021HighDR,LIU2019217} and Moire Technique (MT)\cite{Li:17,CHEN2023109666,Ordones:21}  do not rely on multi-frame projection. At the same time, both of them are sensitive to noise and suffer from spectrum aliasing and leakage when reconstructing objects with strong textures or steep surfaces\cite{XIAO201719,Wang:21,Wang:20}. The single-frame fringe projection 3D measurement method can use machine learning for wrapped phase prediction and phase-shifted fringe prediction. Convolutional neural networks combined with fringe analysis achieve high-precision results in a laboratory environment. These networks are supervised and trained using a large number of simulated or real stripe images and high-precision ground truth. Variants of networks such as FCN\cite{DBLP:journals/corr/LongSD14}, AEN\cite{DBLP:journals/corr/abs-1902-09314}, and U-Net\cite{DBLP:journals/corr/RonnebergerFB15} have been studied as backbone of end-to-end networks, with U-Net being the most popular\cite{Wan:23}. However, the "one-shot" method which captures only one frame of image cannot obtain robust scene information under complex lighting conditions\cite{BURNES2022106788,ODOWD2020106293}. This causes accuracy reduction of 3D reconstruction, especially in dynamic scenes or extremely short exposure conditions.

Binocular vision inherits the advantages of single-frame 3D measurement. With the release of the Scene Flow dataset, DispNet \cite{Mayer_2016} has achieved a breakthrough in the stereo matching field using an encoder-decoder structure. Networks based on the encoder-decoder structure have been developed, as examples of PSMNet \cite{chang2018pyramid} and BGNet \cite{xu2021bilateral}. The stereo matching network based on 3DCNN replicates the classic stereo matching process, including cost calculation and cost aggregation \cite{Wang_2021,shen2022pcwnet,wang2020improving}. Combining classic multi-view geometry and deep learning has repeatedly achieved state-of-the-art results \cite{10203487,chabra2019stereodrnet}. However, 4D cost volume filtering is computationally intensive and consumes enormous memory, making it difficult to generalize to images with millions of pixels \cite{duggal2019deeppruner,xu2020aanet}. Recently, iterative optimization-based methods have been proven to simplify the cost aggregation operation. Yet some inherent problems still need to be addressed: creating typical stereo datasets is labor-intensive and expensive, often limiting the scene scope \cite{Zhao2022EAIStereoEA,tosi2021smdnets,wang2021scvstereo,xu2023iterative,Wang_2024_CVPR}. Trained models perform well in a limited set of scenarios, but without fine-tuning, they perform disappointingly on cross-scenario tasks \cite{zhang2020domain,tonioni2019learning,pang2018zoom}. 

The application of state-of-the-art (SOTA) stereo matching networks to measurements remains an intractable problem. This challenge is further exacerbated when dealing with out-of-domain, low-quality, or perturbed samples. Inspired by phase truncation in structured light techniques, we propose a novel approach that combines Phase Measuring Profilometry \cite{Zheng:17,LI2022106990,Li:16,SONG201674} with stereo matching networks, introducing a color phase speckle projection. By utilizing the PMP-encoded phase speckle map instead of RGB \cite{WU2022107955,KEMAO2022106874} texture as the input for the model, our method enables consistent calculations across various scenes. This approach empowers the model to learn matching relationships and pattern similarities, effectively mitigating obstacles arising from environmental illumination changes and texture discrepancies. When the reflectivity of the object surface changes drastically and the lighting conditions are more stringent, higher requirements are placed on the generalization of the proposed method for reconstruction in the scene. There are many common problems in real scenes, such as environmental color cast, occlusion, uneven reflectance and texture interference. There are many obstacles in the cross-scene reconstruction of the previous stereo matching model using real data sets, which makes it difficult to extend to unseen scenes. As shown in Figure \ref{fig:intro}, our method can achieve better cross-scene reconstruction performance in contrast.

The contributions of this work are summarized as follows:
\begin{itemize}
\item We designed a novel encoding approach by randomly perturbing the phase shift map and integrating it into the three RGB channels to generate a color speckle pattern, which is then projected onto the scene to incorporate prior information.

\item A phase pre-normalization method is proposed to mitigate the interference from environmental lighting and texture while ensuring domain consistency among different input image pairs before entering the stereo matching network. This approach enables stereo matching network trained with this method to exhibit superior reconstruction performance in scenes projected with RGB speckle patterns containing phase information, even for unseen scenes.

\item we constructed a large RGB speckle realistic datasets with sub-pixel precision, capturing realistic and challenging scenes using an active binocular system. We conducted quantitative and qualitative experiments based on this foundation, and the results show that experiments performed on both public datasets and this speckle dataset confirm the robustness and generalizability of our approach.

\end{itemize}
\section{Related Work}
\label{sec:related}

\begin{figure*}[t!]
    \centering
    \includegraphics[width=0.85\textwidth]{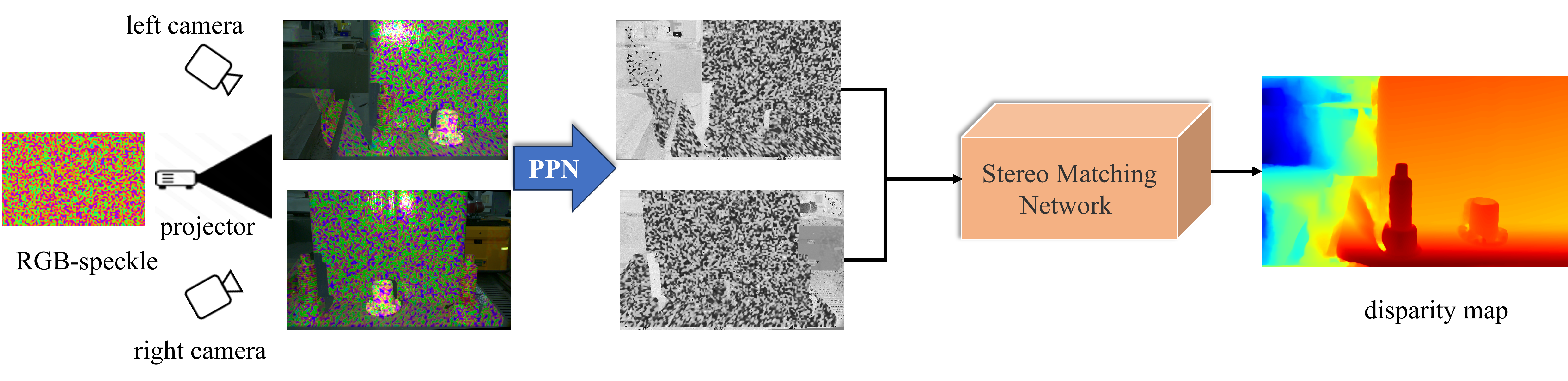} 
    \caption{The pipeline of our method. By projecting phase-encoded color speckle patterns onto the scene during image capture and applying phase pre-normalization, robust reconstruction can be achieved across different scenes using a stereo matching network trained on speckle dataset.}
    \label{fig:pipeline}
    \end{figure*}

\paragraph{Structured light 3D reconstruction}
is considered one of the premier methodologies for optical 3D measurement. In principle, it involves projecting one or several patterns onto the surface of the target \cite{KEMAO2022106874,Feng:15,LU2022106873} and then establishing correspondences between the projector and the camera using these projected motifs. By leveraging the triangulation principle, it becomes feasible to reconstruct the three-dimensional scene through the integration of calibration data with phase unwrapping techniques \cite{Wang:17}. Of them, Fringe Projection Profilometry (FPP) is renowned for its reliability in profiling surfaces \cite{WU2022107004,YU2022128236}, but there are significant obstacles in extracting accurate and steadfast absolute phase information from a single fringe image poses substantial hurdles, limiting its efficacy in dynamic 3D reconstructions \cite{article03}. Speckle Projection Profilometry (SPP) facilitates the establishment of correspondences across stereo images by projecting a singular speckle pattern \cite{article01,Su2019TheoreticalAO}, thus enabling single-shot 3D reconstructions. However, most research primarily focuses on encoding positional data and deciphering correspondences, often overlooking a crucial aspect - minimizing interference caused by ambient light and implicit attributes inherent in the captured image \cite{article02}. 

\paragraph{Stereo matching}
 estimates a pixel-to-pixel map between two frames of images, and typically comprises four main steps: cost computation, cost aggregation, disparity calculation, and disparity refinement\cite{wu2019semantic,zhang2019ganet,tankovich2023hitnet}. In real-world uncertainty conditions (i.e., conditions affected by occlusion, smoothness, reflections, and noise), estimating disparity poses a challenging task\cite{taniai2017continuous,9337529}. With the profound transformation brought by deep learning to computer vision, convolutional neural networks (CNNs) have been introduced into stereo matching pipelines\cite{pang2018cascade,7299044,Seki2016PatchBC}. By exploring different feature representations and aggregation algorithms for matching costs, CNNs can address these challenges\cite{7780983}. For example, Mayer et al. proposed an end-to-end network for estimating disparities  and optical flows  \cite{dosovitskiy2015flownet}, along with a synthetic dataset called Scene Flow, which has become widely used for stereo tasks. In 2017, kenda firstly combined 3DCNN with the classic stereo matching pipeline for cost aggregation\cite{kendall2017endtoend}. Guo et al. proposed a more efficient group-wise correlation volume computation method, capable of better measuring the feature correlation between the left and right images \cite{guo2019groupwise}. Shen et al. adopted a cascade cost volume and adjusted the disparity search range adaptively in subsequent processes based on variance uncertainty estimation  \cite{shen2021cfnet}. Xu et al. obtained attention weights from correlation features to suppress redundant information in the concatenated volume and enhance its expressive capability \cite{xu2022attention}. Chen et al.optimize feature channels through motif channel projection instead of direct network optimization to mitigate imbalanced nonlinear transformations in network training\cite{Chen_2024_CVPR}. Gong et al. learn intra-view geometric and cross-view knowledge for stereo matching networks from an interest point detector and an an interest point matcher\cite{Gong_2024_CVPR}. End-to-end network structures dominated by 3D CNNs exhibit satisfactory performance and even produce high-quality results in pathological areas on ideal disparity maps of public datasets\cite{7410474,du2019amnet,song2021adastereo}. Although these models performed well in the training phase on specific datasets, their ability to generalize to natural scenes has not been validated. Moreover, the model heavily relies on a large amount of high-quality ground truth, it remains unexplored how to make the binocular matching model perform good results on unseen datasets.

It is the focus of cross-domain stereo matching research to be unaffected by pre-offset between different data sets. In this regard, Zhang et al. proposed a new domain normalization method, which normalizes the distribution of learning representations and makes them invariant to domain differences\cite{zhang2020domain}. Chuah W Q et al. proposed an information theory shortcut avoidance method, the information related to the shortcut is automatically limited to be encoded into the feature representation\cite{chuah2022itsa}. Pang et al. proposed a semi-supervised method using scale information\cite{pang2018zoom}. Guo et al. proposed a cross-domain method using knowledge extraction\cite{guo2018learning}. Recently, Zhang et al. avoided domain generalization ability degraded from synthetic to real by dividing pixels into consistent and inconsistent regions based on the difference between ground truth and pseudo label\cite{zhang2024robust}. However, these methods do not implement standard and complete domain adaptation channels, and their adaptation performance is very limited. Differert from these methods, we propose a robust RGB-Phase Speckle method based on pseudo-random arrays, which improve the generalizability of the algorithm by making it insensitive to ambient lighting using phase feature constraints.

Different from these methods, by introducing additional prior information and normalizing the scene, our approach bypasses the challenges of cross-domain generalization faced by stereo matching networks. It ensures domain consistency across input data, thereby preventing the network from overfitting to domain-specific datasets during training.

\section{Proposed Method}

This section outlines the methodological architecture of the RGB-Phase Speckle approach. Starting with presenting the RGB-Phase Speckle formula, we introduce the underlying principles that bolster its efficacy directly from the formulation. Then, we introduced the color speckle dataset under complex scenes produced according to the proposed method, which can be used for training stereo matching networks under cross-scene applications. The pipeline of our method is shown in Figure \ref{fig:pipeline}.

\subsection{RGB-Phase Speckle}

The phase information can reflect the height change of the surface of the object. Inspired by color phase-shifted fringe patterns based on phase measuring profilometry\cite{LIU2020106267}, our patterns are composed from three phase-shifted speckles encoded in the RGB channels.

\begin{figure}
    \centering
    \includegraphics[width=1\linewidth]{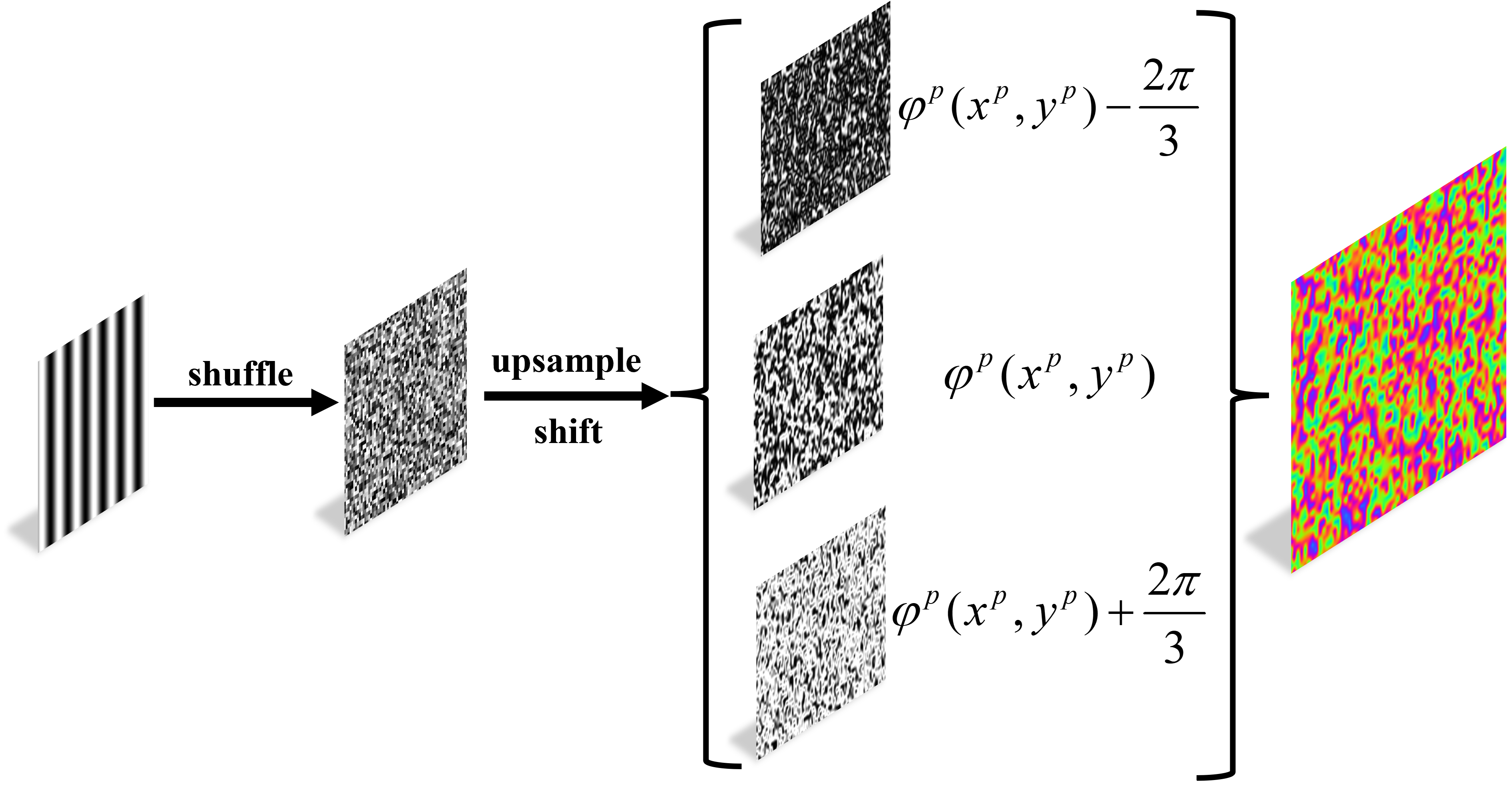}
    \caption{The generation process of phase color speckle pattern is disturbed and upsampled in the same way on a phase-shifting fringe and merged into three RGB channels to realize the phase-shifting coding of a single image.}
    \label{fig:rgb}
\end{figure}

To mitigate the impact of an object's inherent texture on color fringe information, we apply identical scrambling to the three-step phase-shifting fringes to generate the color speckle pattern. This approach ensures that the phase modulation at each point on the object remains independent of the relative positional relationships between the fringes, thereby facilitating network training. Furthermore, upsampling from a low-resolution speckle image prevents the speckle pattern from being submerged in sensor noise. The speckle generation process is illustrated in Figure\ref{fig:rgb}.

Taking the red channel pattern in the projector as an example, \((x^p,y^p)\) represents the pixel coordinates of the projected pattern, \(a^p\) denotes the background intensity, and \(b^p\) represents the amplitude,  \(\varphi^p(x^p,y^p)\) represents the encoded phase, \(G^p\left(x^p, y^p\right)\) is the superposition of the phase-related pattern. The RGB channels have a phase difference of \( \frac{2\pi}{3} \) to each other. The combined colors of the blue channel \(B^p(x^p, y^p)\), green channel \(G^p(x^p, y^p)\), and red channel \(G^p(x^p, y^p)\) are as follows:
\begin{align}
    B^p\left(x^p, y^p\right)=a^p+b^p \cos \left[\varphi^p(x^p,y^p) - \frac{2\pi}{3} \right]
\end{align}
\begin{align}
    G^p\left(x^p, y^p\right)=a^p+b^p \cos \left[\varphi^p(x^p,y^p)\right]
\end{align}
\begin{align}
    R^p\left(x^p, y^p\right)=a^p+b^p \cos \left[ \varphi^p(x^p,y^p) +\frac{2\pi}{3} \right]
\end{align}

It is noteworthy that the single CMOS color camera employ a Bayer array color filter placed in front of the sensor with a demosaicing algorithm used to interpolate the RGB values of pixels. Due to manufacturing imperfections, the passband ranges of the color filters for different channels partially overlap. Light reflections at wavelengths within these overlapping regions induce crosstalk among pixels across different channels, resulting in shifts in interpolation-estimated pixel values and introducing a phase deviation, denoted as $\Delta\varphi$. This deviation constitutes a non-negligible error in structured light systems relying on single-camera projection. In binocular systems, where the phase deviation $\Delta\varphi$ is consistent across both cameras, our projection pattern has no adverse effect on stereo matching performance. Thus, our approach requires no strict consistency in the projected color speckle patterns, such as eliminating the need for precise three-channel alignment.

\subsection{Phase Pre-Normalization}

Our RGB phase speckle patterns introduce prior information into the scene. However, this phase information is embedded within the color speckle texture projected onto the object. Variations in reflectance and illumination conditions across objects in different scenes, coupled with inconsistencies in the image signal processing (ISP) algorithms of different cameras, result in domain shifts in the captured image pairs. These inconsistencies contribute significantly to the limited cross-domain generalization performance of stereo matching networks. To overcome this challenge, we introduce a phase pre-normalization (PPN) method:
\begin{align}
    \varphi(x,y) &= \begin{aligned}[t]
    &\tan^{-1}\frac{2G(x,y)-R(x,y)-B(x,y)}{R(x,y)-B(x,y)} \\
    &= \tan^{-1}\frac{A(2g-b-r)+B\cos\varphi(2g+\frac{b}{2}+\frac{r}{2})}{A(r-b)+B\cos\varphi(\frac{r}{2}+\frac{b}{2})} \\
    &= \tan^{-1}\frac{B\cos\varphi(2g+\frac{b}{2}+\frac{r}{2})+w_1}{B\cos\varphi(\frac{r}{2}+\frac{b}{2})+w_2}
    \label{eq:norm}
    \end{aligned}
    \end{align}

\begin{figure}
    \centering
    \includegraphics[width=0.85\linewidth]{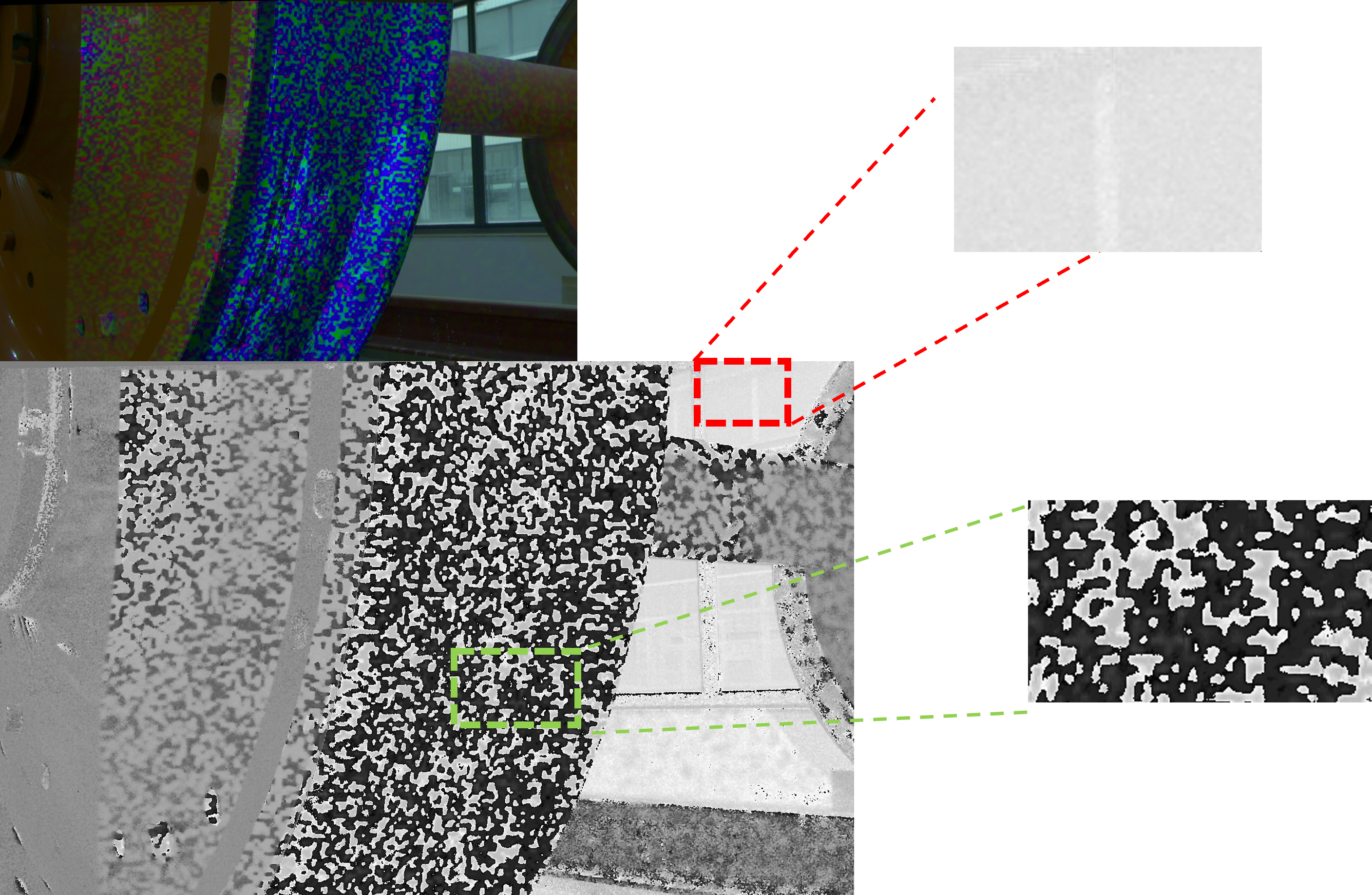}
    \caption{In the scene of projected color speckle and the normalized image, it can be seen that the area projected speckle retains the phase information after pre-normalization, while the background area not projected speckle is ignored in this process.}
    \label{fig:norm}
\end{figure}

Similar to the wrapped phase, the phase of deformed speckle is also constrained within \((- \pi, \pi)\). Equation \ref{eq:norm} indicates that, with an additional phase computation constraint, background light intensity is attenuated, while the active projection is amplified. Specifically, when the coefficients  \(r\), \(g\), \(b\) are nearly equal, the terms \(w_1\) and \(w_2\) become negligible, which means utilizing normalized phase as model input enhances feature distribution consistency across datasets. The pre-normalized phase is shown in Figure \ref{fig:norm}.

It can be observed that the phase information of the region within the green box, where speckle patterns are projected, is effectively preserved after pre-normalization. In contrast, the background region within the red box, which lacks projected colored speckles, has its texture information eliminated following pre-normalization. Consequently, in the image pairs input to the network, only the phase information from the object's surface is retained. This approach mitigates the adverse effects of ambient illumination and irrelevant background regions on the matching process. Moreover, the pre-normalization process leverages the Color Correction Matrix (CCM) characteristics of the color camera, ensuring that phase information is preserved in normally exposed images. This hardware-level consistency fundamentally guarantees domain uniformity, thereby alleviating issues related to inconsistent feature distributions across different scenes.
\subsection{RGB-Speckle Datasets}

\begin{figure} 
    \centering
    
    \begin{subfigure}[b]{0.45\linewidth} 
        \centering
        \includegraphics[width=\linewidth]{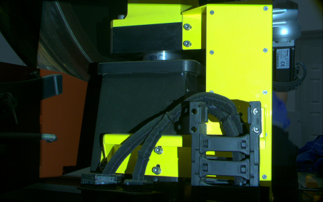} 
        \caption{Origin scene}
        \label{fig:industrial scene}
    \end{subfigure}
    \hspace{2mm}
    \begin{subfigure}[b]{0.45\linewidth}
        \centering
        \includegraphics[width=\linewidth]{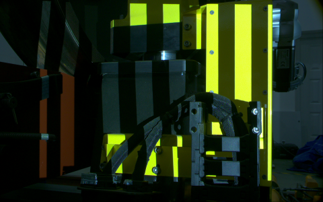}
        \caption{Graycode}
        \label{fig:graycode}
    \end{subfigure}
    \begin{subfigure}[b]{0.45\linewidth}
        \centering
        \includegraphics[width=\linewidth]{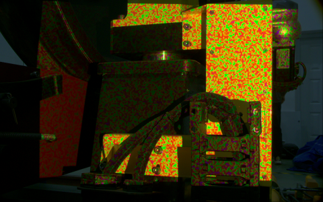}
        \caption{Speckle}
        \label{fig:subfig3}
    \end{subfigure}
    \hspace{2mm}
    \begin{subfigure}[b]{0.45\linewidth}
        \centering
        \includegraphics[width=\linewidth]{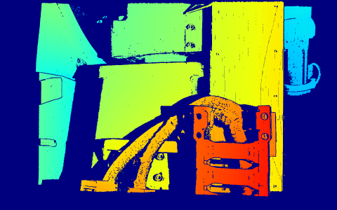}
        \caption{ Ground truth }
        \label{fig:subfig4}
    \end{subfigure}
    \caption{The production process of our speckle dataset. (a) A typical scene of our speckle dataset. (b)One of projected Gray code fringe. (c) Our projected color speckle. (d) The sub-pixel disparity ground truth.}
    \label{fig:dataset}
\end{figure}

Since existing public stereo datasets lack prior information, training the network on texture-based image pairs limits the ability to effectively learn phase features. Therefore, we employ an active binocular system to capture  complicated scenes projected by our RGB speckle patterns and the dense subpixel disparity ground truth by projecting Gray code patterns\cite{li2024three}. The scenes of our speckle datasets have abundant differences surface reflectance objects and repeated textures, which is very challenging for stereo matching. The illustration of our dataset is shown in Figure\ref{fig:dataset}.

Though the ill-posed areas in these scenes are textured by speckles, it is difficult to obtain better results for the stereo matching network without the training of this kind of data. The reason is that the traditional method relies on color texture features, which shows poor generalization for scenes with huge differences in object surface reflectance, and the data of the network in other scenes will be affected by serious domain offset.

On the contrary, through the phase domain normalization and the phase information of the active projection, the model learns the matching relationship between the phase information independent of the reflectivity of the object surface, thus avoiding the domain offset problem, so that the model can still get better results in other scenarios.

 By fine-tuning the stereo matching network with our phase pre-normalized speckle dataset,images captured by active binocular system that can project color speckles can reconstructed robustly even in previously unseen scenes. We anticipate that our provided speckle dataset will contribute to advancing the application and development of deep learning-based active binocular systems.

\section{Experiments}

\begin{table*}
    \centering
    \begin{tabular}{cccc}
    \toprule
         Training & Validation & EPE & D1\\
    \midrule
        Scene Flow & Speckle (\textbf{PPN})& 21.406986 & 15.692858\\
        Scene Flow & Speckle & 15.853035 & 9.877504\\
        Scene Flow (\textbf{PPN}) & Speckle & 8.331100 & 5.238395\\
        Scene Flow (\textbf{PPN}) & Speckle (\textbf{PPN})& 5.120400 & 5.496406\\
        Speckle (\textbf{PPN}) & Speckle & 1.074344& 1.948889\\
        Speckle (\textbf{PPN}) & Speckle (\textbf{PPN})& \textbf{0.816987}& \textbf{1.105160}\\
    \bottomrule
    \end{tabular} 
    \caption{Comparison of ablation experiment  with EPE and D1}
    \label{tab:my_label} 
\end{table*}

\begin{figure*}
    \centering
    \includegraphics[width=0.85\linewidth]{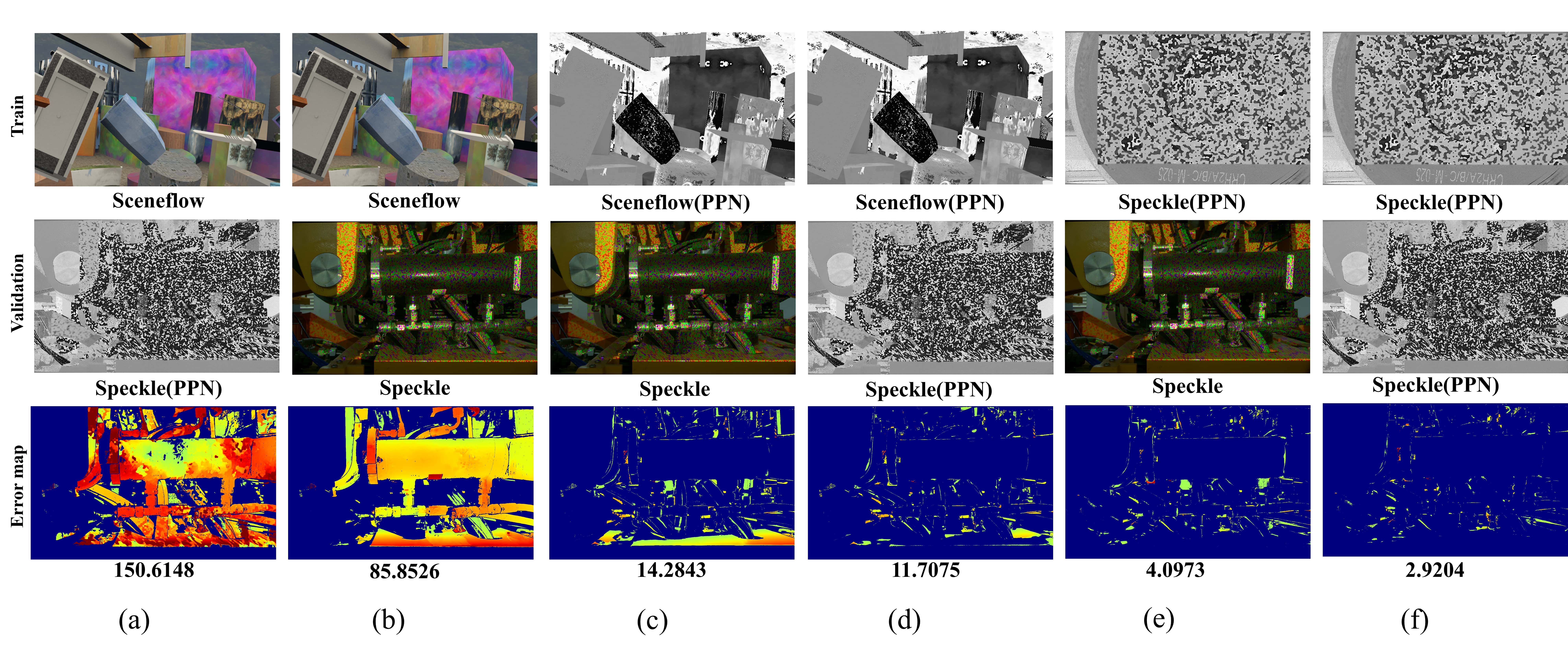}
    \caption{Accuracy performance across different training strategies. (a) Training on sceneflow and validation on phase pre-normalized(PPN) Speckle. (b) Training on sceneflow and validation on RGB Speckle. (c)Training on PPN sceneflow and validation on RGB Speckle. (d) Training on PPN sceneflow and validation on PPN Speckle. (e) Training on PPN Speckle and validation on RGB Speckle. (f) Training on PPN Speckle and validation on PPN Speckle. From top to bottom, the rows represent the input during the training phase, the input during the inference phase, and the error distribution within the mask. The bottom row shows the End-Point Error (EPE) of the computed disparity map.
    }
    \label{fig:scene_gt}
\end{figure*}

To validate the robustness of our approach, we conducted comprehensive quantitative and qualitative experiments. First, through ablation study, we analyzed the contribution of our phase-based method to the generalization performance of the network. Furthermore, we evaluated the robustness of our proposed method by comparing its performance across different stereo matching networks and various speckle patterns. Finally, we performed 3D reconstruction on unseen scenes to assess the method's cross-domain performance.

\subsection{Datasets and Metrics}
\paragraph{Datasets}To assess the performance of our proposed models and benchmark them against state-of-the-art methods, we evaluate the depth estimation accuracy using the SceneFlow dataset\cite{mayer2016large} and our custom RGB speckle dataset. The SceneFlow dataset, a widely recognized large-scale synthetic stereo matching benchmark, comprises 35,454 stereo image pairs for training and 4,370 pairs for testing, each with a resolution of 960$\times$540 pixels. Additionally, our RGB speckle dataset is designed to complement this evaluation by incorporating real-world speckle-patterned imagery, enabling a comprehensive analysis of depth estimation under distinct scene domains. And our speckle scene dataset contains 10716 image pairs and the resolution of the images is 1920$\times$1200 pixels acquired by an active binocular system. Given that our speckle dataset exhibits a significantly wider disparity range, we conducted training on the SceneFlow dataset at double the resolution to mitigate the risk of training instability or non-convergence caused by excessive disparities between the datasets.

\paragraph{Metrics}Consistent with prior work in stereo matching, we adopt End-Point Error (\textbf{EPE}) and \textbf{D1} error rate as evaluation metrics to assess the performance of our proposed method. EPE is defined as the average absolute disparity error computed pixel-wise between the predicted disparity map and the corresponding ground truth. Meanwhile, the D1 error rate quantifies the fraction of pixels for which the absolute disparity error surpasses a specified threshold, such as 3 pixels.

\subsection{Implementation Details}


The experiments were conducted on four NVIDIA RTX 3090 GPUs. For the ablation study, we selected IGEV as the baseline model\cite{xu2023iterative}. The training process consisted of 140,000 iterations with a batch size of 12. Subsequently, we fine-tuned the model on our Speckle dataset for 350,000 iterations, employing a dataset split of 6:2:2 for training, validation, and testing, respectively. The initial learning rate was set to 0.0002, and training was performed using PyTorch with the AdamW optimizer, following the One-Cycle learning rate scheduling strategy. 

In stereo matching comparison experiments, we evaluate our approach against PSM-Net, ACV-Net, and RAFT-Stereo, three representative stereo matching networks\cite{chang2018pyramid,xu2022attention,lipson2021raftstereo}. Each network was trained to convergence on both the SceneFlow and Speckle datasets using PPN method.

We employed a binocular vision systems system consists of a color projector (XGIMI Z6X, resolution 1280$\times$720 pixels) paired with two cameras (Basler ace acA1920-40gc, resolution 1920$\times$1200 pixels, focal length 12 mm), featuring a baseline separation of 165 mm. The working distance for our system spans 0.5 to 1.5 meters.

\subsection{Ablation Study}

Since our method relies on an active binocular system to project additional speckle information onto objects, a direct comparison with prior approaches trained on public datasets would be inherently biased. Consequently, we designed our ablation studies to leverage both the SceneFlow and Speckle datasets. Specifically, we employed distinct training strategies on these datasets by incorporating phase pre-normalization selectively during training. The final evaluation was conducted exclusively on the validation set of our Speckle dataset, where we computed the End-Point Error (EPE) and D1 error rate to validate the enhancement of our method in improving the generalization performance of networks across diverse scenes.

To rigorously assess the effectiveness of our proposed method, we systematically partitioned the training and validation process into six controlled experimental groups. The performance of these groups is illustrated in Figure\ref{fig:scene_gt}:

Firstly, We conducted inference on the speckle dataset using the weights pretrained on the SceneFlow dataset. The results indicate a markedly poor generalization performance, predominantly due to the network's dependence on color and texture features extracted directly from RGB images during the training process, especially on the speckle data after phase pre-normalization(PPN), where the significant domain gap renders the weights trained on RGB images entirely incompatible with phase maps, as illustrated in Figure\ref{fig:scene_gt} a. Moreover, a network trained solely on color and texture features exhibits domain sensitivity in scenes characterized by significant variations in object surface reflectance. Consequently, inference performance in other scenes is severely impacted by substantial domain shift, as demonstrated in Figure\ref{fig:scene_gt} b.

For the second training strategy, we apply Phase Pre-Normalization (PPN) to the SceneFlow dataset. Following network training on the PPN-processed iamge pairs, the accuracy on our speckle dataset exhibits greatly improvement. It demonstrate the effectiveness of our approach in enhancing cross-domain performance from synthetic to real datasets, as illustrated in Figure\ref{fig:scene_gt} c. Furthermore, when we inference on the speckle dataset after PPN, the network achieves higher accuracy, with notably lower error compared to the non-PPN-processed data, as shown in Figure Figure\ref{fig:scene_gt} d. This observation suggests that, even across datasets ranging from synthetic to real or from non-speckled to speckled domains, the normalized data domains remain closely aligned.

Finally, we apply PPN on the speckle dataset and perform inference on both the original speckle images and PPN images respectively. The error is much lower than other group, as shown in \ref{fig:scene_gt}e and \ref{fig:scene_gt}f, and the best performance is achieved on the images after PPN, which fully illustrates the cross-scene capability of our proposed method.

Table \ref{tab:my_label} shows the indicators calculated by our six training methods on the validation set of our speckle images. From the perspective of EPE and D1, the IGEV network trained with pre-normalization phase images has the best performance on the same PPN speckle validation images.Even in the speckle data without PPN, it can also have excellent performance. This result clearly demonstrates that a network trained on abundant normalized phase data effectively overcomes its reliance on color and texture features, enabling it to capture deeper phase-related features. Consequently, it delivers superior performance on images containing phase information without the need for pre-normalization, a capability that remains difficult to achieve across datasets devoid of phase information.

\begin{figure} 
    \centering
    
    \begin{subfigure}[b]{1\linewidth} 
        \centering
        \includegraphics[width=\linewidth]{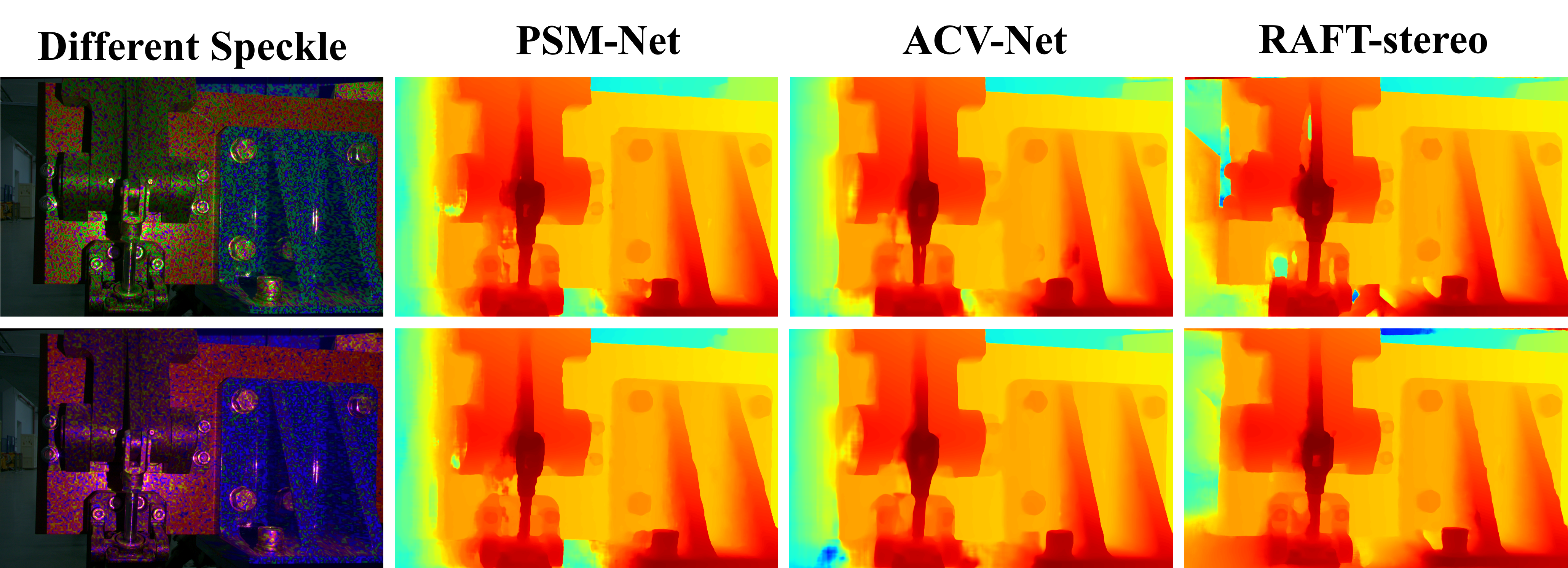} 
        \caption{Scene 1}
        \label{fig:scene1}
    \end{subfigure}
    
    \begin{subfigure}[b]{1\linewidth}
        \centering
        \includegraphics[width=\linewidth]{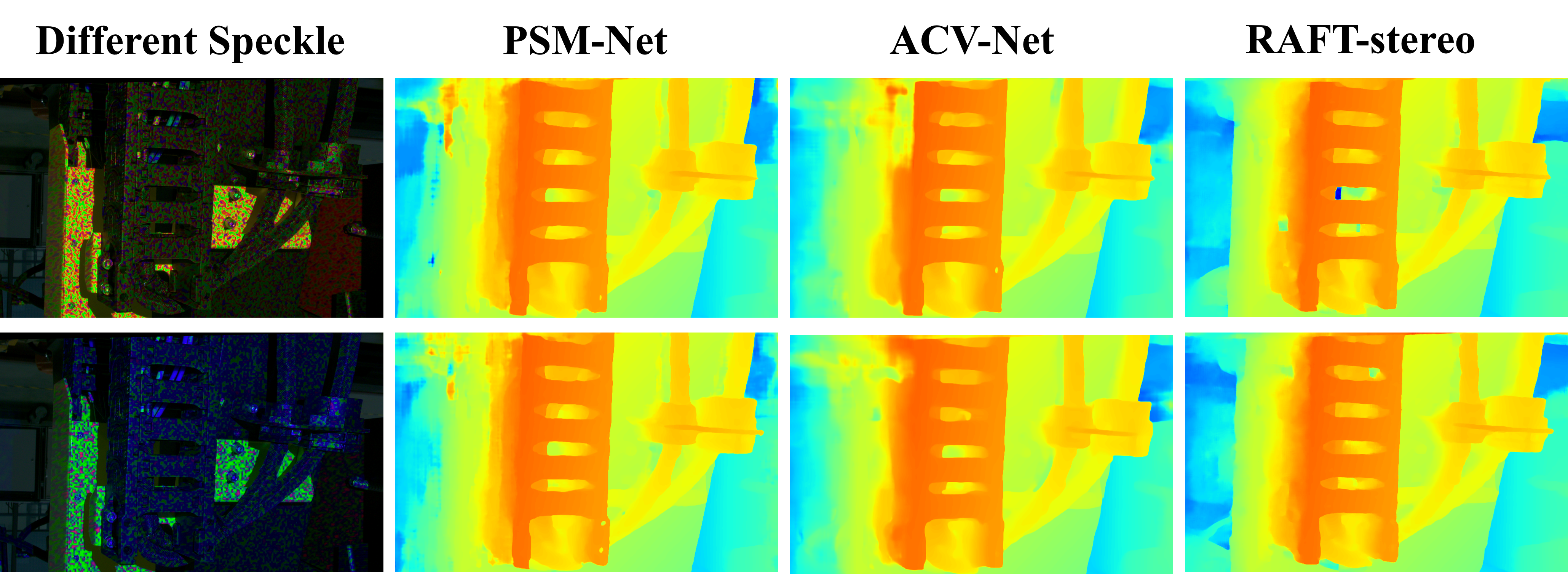} 
        \caption{Scene 2}
        \label{fig:scene2}
    \end{subfigure}
    
    \caption{The performance of different stereo matching networks under different projection speckle patterns.}
    \label{fig:net}
\end{figure}

Additionally, our approach imposes no constraints on the choice of stereo matching networks. The fundamental reason lies in the fact that we do not enhance cross-domain performance by modifying the network architecture itself; rather, we address this from the data perspective, improving the domain alignment of input data through preprocessing. To substantiate this, we trained PSM-Net, ACV-Net, and RAFT-Stereo on speckle data processed with PPN and conducted inference on two selected scenes from the speckle validation set with PPN, as illustrated in Figure\ref{fig:net}.

    As observed in the figure, these networks consistently achieve high-precision reconstruction results in regions containing phase information, underscoring the robustness of our approach. Additionally, we explored various permutations of speckle patterns disrupted in different ways across channels. The resulting new speckle patterns also yield favorable outcomes, attributable to the fact that phase information, modulated by the object, is embedded within a single image. Swapping this information across the three channels merely shifts the decoded phase values within the $(-\pi, \pi)$ interval without incurring any loss. Even when speckle patterns are processed through projection devices and the camera's Bayer filter, introducing phase deviations, the binocular system leverages redundant information to enhance robustness against noise, thereby minimizing adverse effects on disparity map reconstruction. Our method remains effective irrespective of system variations, provided the disparity range during training and inference remains consistent, rendering it applicable to other active binocular systems.

%

\begin{figure}
    \centering
    \includegraphics[width=1\linewidth]{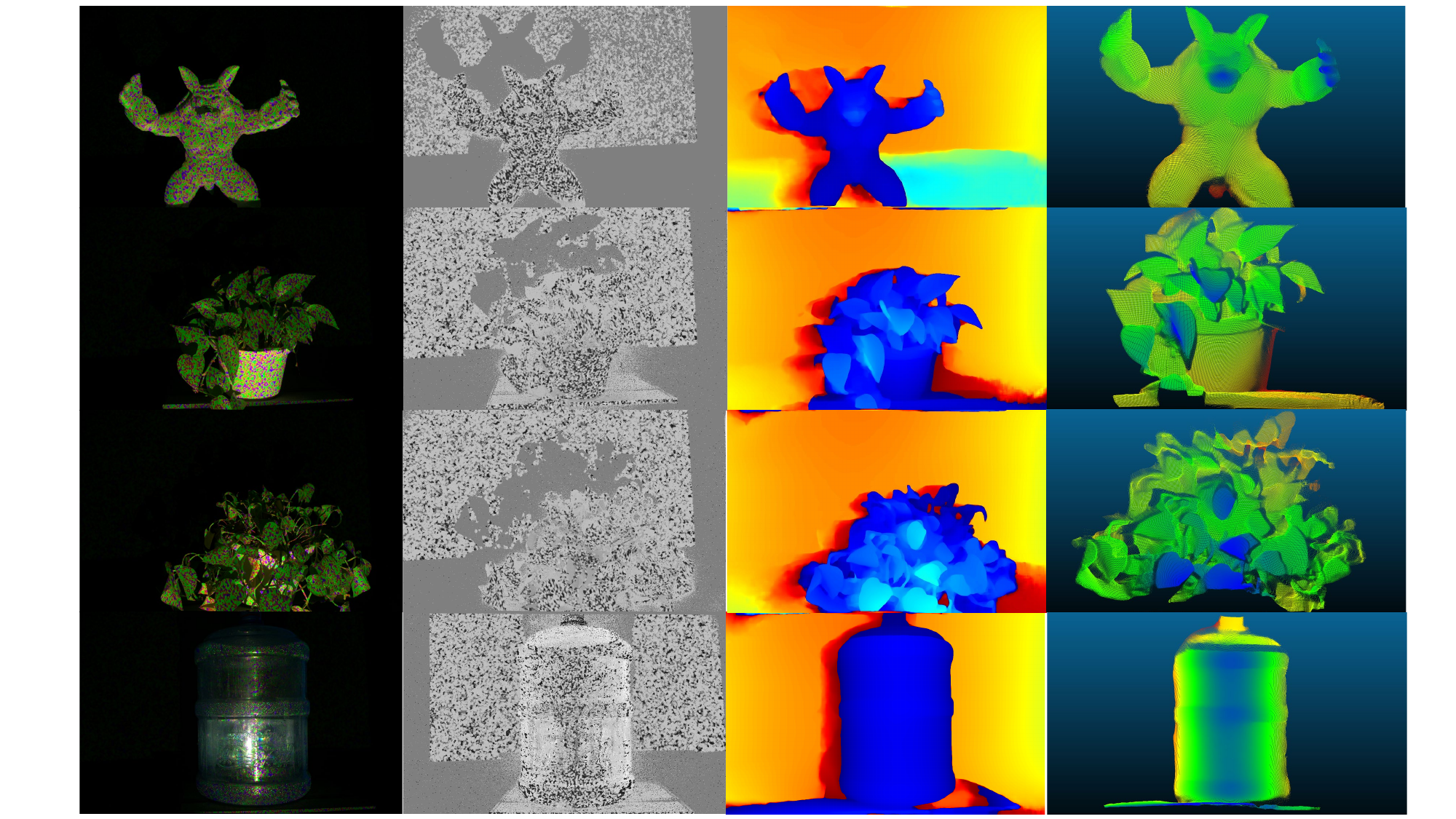}
    \caption{Cross-scene performance display. The reconstruction details can be seen from the rightmost point cloud. }
    \label{fig:cross-scene}
\end{figure}

The performance in unseen scenarios is our primary concern, as it directly determines whether our approach can be effectively applied to real-world 3D perception and precision measurement tasks. To rigorously evaluate generalization, we select scenarios that are entirely different from the synthetic and speckle data used for training. By projecting RGB speckle patterns and following our inference pipeline, we obtain the results shown in Figure \ref{fig:cross-scene}.

The figure presents various objects, including a monster model, two potted plants and a bucket. Our method demonstrates strong performance in these unseen scenarios. The reconstructed point cloud, derived from the disparity map, reveals that even in previously unseen regions—particularly along object boundaries—fine details are well preserved. This is attributed to the consistency of the phase information embedded in the speckle pattern across non-occluded areas. The experimental results validate the strong cross-scene generalization capability of our approach, offering a novel solution for active stereo systems.


\section{Conclusion}
In this study, we proposed a novel RGB-Phase Speckle approach that introduces three key innovations: the generation of RGB color phase speckles, a phase pre-normalization technique, and a new speckle dataset tailored for stereo matching tasks. Our method effectively addresses the significant reliance of deep learning-based active binocular systems on data consistency by leveraging phase pre-normalization, which aligns data domains and enhances model generalization. This advancement facilitates the practical deployment of active binocular systems in diverse real-world scenarios. Experimental results demonstrate that our method achieves accurate and robust depth inference across images captured by physical systems with identical configurations, irrespective of the scene. This improved generalization highlights the potential of our approach in challenging environments where traditional methods struggle.

Future research will focus on optimizing projection hardware to achieve reduced costs, smaller form factors, faster projection speeds, and improved energy efficiency. Additionally, we aim to further explore the miniaturization, integration, and generalization of active binocular systems to expand their practical applicability in real-world measurement tasks.

{
    \small
    \bibliographystyle{ieeenat_fullname}
    \bibliography{main}

\begin{thebibliography}{76}
\providecommand{\natexlab}[1]{#1}
\providecommand{\url}[1]{\texttt{#1}}
\expandafter\ifx\csname urlstyle\endcsname\relax
  \providecommand{\doi}[1]{doi: #1}\else
  \providecommand{\doi}{doi: \begingroup \urlstyle{rm}\Url}\fi

\bibitem[Bai et~al.(2020)Bai, Yang, Li, and Sui]{9337529}
Zi-Jian Bai, Kai Yang, Jin-Long Li, and Hao Sui.
\newblock Occlusion area removal in binocular 3d reconstruction of train running parts.
\newblock In \emph{2020 IEEE Far East NDT New Technology \& Application Forum (FENDT)}, pages 155--159, 2020.

\bibitem[Burnes et~al.(2022)Burnes, Villa, Moreno, {de la Rosa}, Alaniz, and González]{BURNES2022106788}
Susana Burnes, Jesús Villa, Gamaliel Moreno, Ismael {de la Rosa}, Daniel Alaniz, and Efrén González.
\newblock Temporal fringe projection profilometry: Modified fringe-frequency range for error reduction.
\newblock \emph{Optics and Lasers in Engineering}, 149:\penalty0 106788, 2022.

\bibitem[Chabra et~al.(2019)Chabra, Straub, Sweeney, Newcombe, and Fuchs]{chabra2019stereodrnet}
Rohan Chabra, Julian Straub, Chris Sweeney, Richard Newcombe, and Henry Fuchs.
\newblock Stereodrnet: Dilated residual stereo net, 2019.

\bibitem[Chang and Chen(2018)]{chang2018pyramid}
Jia-Ren Chang and Yong-Sheng Chen.
\newblock Pyramid stereo matching network, 2018.

\bibitem[Chen et~al.(2023)Chen, Zhang, Shi, and Xie]{CHEN2023109666}
Ru Chen, ChengHao Zhang, Wenxiong Shi, and Huimin Xie.
\newblock 3d sampling moiré measurement for shape and deformation based on the binocular vision.
\newblock \emph{Optics \& Laser Technology}, 167:\penalty0 109666, 2023.

\bibitem[Chen et~al.(2015)Chen, Sun, Wang, Yu, and Huang]{7410474}
Zhuoyuan Chen, Xun Sun, Liang Wang, Yinan Yu, and Chang Huang.
\newblock A deep visual correspondence embedding model for stereo matching costs.
\newblock In \emph{2015 IEEE International Conference on Computer Vision (ICCV)}, pages 972--980, 2015.

\bibitem[Chen et~al.(2024)Chen, Long, Yao, Zhang, Wang, Qin, and Wu]{Chen_2024_CVPR}
Ziyang Chen, Wei Long, He Yao, Yongjun Zhang, Bingshu Wang, Yongbin Qin, and Jia Wu.
\newblock Mocha-stereo: Motif channel attention network for stereo matching.
\newblock In \emph{Proceedings of the IEEE/CVF Conference on Computer Vision and Pattern Recognition (CVPR)}, pages 27768--27777, 2024.

\bibitem[Chuah et~al.(2022)Chuah, Tennakoon, Hoseinnezhad, Bab-Hadiashar, and Suter]{chuah2022itsa}
WeiQin Chuah, Ruwan Tennakoon, Reza Hoseinnezhad, Alireza Bab-Hadiashar, and David Suter.
\newblock Itsa: An information-theoretic approach to automatic shortcut avoidance and domain generalization in stereo matching networks.
\newblock In \emph{Proceedings of the IEEE/CVF Conference on Computer Vision and Pattern Recognition}, pages 13022--13032, 2022.

\bibitem[Dosovitskiy et~al.(2015)Dosovitskiy, Fischer, Ilg, Hausser, Hazirbas, Golkov, Van Der~Smagt, Cremers, and Brox]{dosovitskiy2015flownet}
Alexey Dosovitskiy, Philipp Fischer, Eddy Ilg, Philip Hausser, Caner Hazirbas, Vladimir Golkov, Patrick Van Der~Smagt, Daniel Cremers, and Thomas Brox.
\newblock Flownet: Learning optical flow with convolutional networks.
\newblock In \emph{Proceedings of the IEEE international conference on computer vision}, pages 2758--2766, 2015.

\bibitem[Du et~al.(2019)Du, El-Khamy, and Lee]{du2019amnet}
Xianzhi Du, Mostafa El-Khamy, and Jungwon Lee.
\newblock Amnet: Deep atrous multiscale stereo disparity estimation networks, 2019.

\bibitem[Duggal et~al.(2019)Duggal, Wang, Ma, Hu, and Urtasun]{duggal2019deeppruner}
Shivam Duggal, Shenlong Wang, Wei-Chiu Ma, Rui Hu, and Raquel Urtasun.
\newblock Deeppruner: Learning efficient stereo matching via differentiable patchmatch, 2019.

\bibitem[Feng et~al.(2015)Feng, Chen, and Zuo]{Feng:15}
Shijie Feng, Qian Chen, and Chao Zuo.
\newblock Graphics processing unit\&\#x2013;assisted real-time three-dimensional measurement using speckle-embedded fringe.
\newblock \emph{Appl. Opt.}, 54\penalty0 (22):\penalty0 6865--6873, 2015.

\bibitem[Gong et~al.(2024)Gong, Liu, Gu, Yang, and Cheng]{Gong_2024_CVPR}
Rui Gong, Weide Liu, Zaiwang Gu, Xulei Yang, and Jun Cheng.
\newblock Learning intra-view and cross-view geometric knowledge for stereo matching.
\newblock In \emph{Proceedings of the IEEE/CVF Conference on Computer Vision and Pattern Recognition (CVPR)}, pages 20752--20762, 2024.

\bibitem[Guo et~al.(2018)Guo, Li, Yi, Ren, and Wang]{guo2018learning}
Xiaoyang Guo, Hongsheng Li, Shuai Yi, Jimmy Ren, and Xiaogang Wang.
\newblock Learning monocular depth by distilling cross-domain stereo networks.
\newblock In \emph{Proceedings of the European conference on computer vision (ECCV)}, pages 484--500, 2018.

\bibitem[Guo et~al.(2019)Guo, Yang, Yang, Wang, and Li]{guo2019groupwise}
Xiaoyang Guo, Kai Yang, Wukui Yang, Xiaogang Wang, and Hongsheng Li.
\newblock Group-wise correlation stereo network, 2019.

\bibitem[Güney and Geiger(2015)]{7299044}
Fatma Güney and Andreas Geiger.
\newblock Displets: Resolving stereo ambiguities using object knowledge.
\newblock In \emph{2015 IEEE Conference on Computer Vision and Pattern Recognition (CVPR)}, pages 4165--4175, 2015.

\bibitem[Kemao(2022)]{KEMAO2022106874}
Qian Kemao.
\newblock Carrier fringe pattern analysis: Links between methods.
\newblock \emph{Optics and Lasers in Engineering}, 150:\penalty0 106874, 2022.

\bibitem[Kendall et~al.(2017)Kendall, Martirosyan, Dasgupta, Henry, Kennedy, Bachrach, and Bry]{kendall2017endtoend}
Alex Kendall, Hayk Martirosyan, Saumitro Dasgupta, Peter Henry, Ryan Kennedy, Abraham Bachrach, and Adam Bry.
\newblock End-to-end learning of geometry and context for deep stereo regression, 2017.

\bibitem[Li et~al.(2016)Li, Liu, and Zhang]{Li:16}
Beiwen Li, Ziping Liu, and Song Zhang.
\newblock Motion-induced error reduction by combining fourier transform profilometry with phase-shifting profilometry.
\newblock \emph{Opt. Express}, 24\penalty0 (20):\penalty0 23289--23303, 2016.

\bibitem[Li et~al.(2017)Li, Cao, Chen, Wan, Fu, and Wang]{Li:17}
Chengmeng Li, Yiping Cao, Cheng Chen, Yingying Wan, Guangkai Fu, and Yapin Wang.
\newblock Computer-generated moir\&\#x00e9; profilometry.
\newblock \emph{Opt. Express}, 25\penalty0 (22):\penalty0 26815--26824, 2017.

\bibitem[Li et~al.(2022)Li, Cao, Wan, Xu, Zhang, An, and haitao Wu]{LI2022106990}
Hongmei Li, Yiping Cao, Yingying Wan, Cai Xu, Hechen Zhang, Haihua An, and haitao Wu.
\newblock An improved temporal phase unwrapping based on super-grayscale multi-frequency grating projection.
\newblock \emph{Optics and Lasers in Engineering}, 153:\penalty0 106990, 2022.

\bibitem[Li et~al.(2024)Li, Yang, Wan, Bai, Liu, Wang, and Xie]{li2024three}
Xinyu Li, Kai Yang, Yingying Wan, Zijian Bai, Yunxuan Liu, Yong Wang, and Liming Xie.
\newblock Three-dimensional reconstruction based on binocular structured light with an error point filtering strategy.
\newblock \emph{Optical Engineering}, 63\penalty0 (3):\penalty0 034102--034102, 2024.

\bibitem[Lipson et~al.(2021)Lipson, Teed, and Deng]{lipson2021raftstereo}
Lahav Lipson, Zachary Teed, and Jia Deng.
\newblock Raft-stereo: Multilevel recurrent field transforms for stereo matching, 2021.

\bibitem[Liu and Kofman(2019)]{LIU2019217}
Xinran Liu and Jonathan Kofman.
\newblock Real-time 3d surface-shape measurement using background-modulated modified fourier transform profilometry with geometry-constraint.
\newblock \emph{Optics and Lasers in Engineering}, 115:\penalty0 217--224, 2019.

\bibitem[Liu et~al.(2020)Liu, Yu, Xue, Zhang, and Su]{LIU2020106267}
Yuankun Liu, Xin Yu, Junpeng Xue, Qican Zhang, and Xianyu Su.
\newblock A flexible phase error compensation method based on probability distribution functions in phase measuring profilometry.
\newblock \emph{Optics \& Laser Technology}, 129:\penalty0 106267, 2020.

\bibitem[Liu et~al.(2021)Liu, Fu, Zhuan, Zhong, and Guan]{Liu2021HighDR}
Yanzhao Liu, Yanjun Fu, Yuhao Zhuan, Kejun Zhong, and Bingliang Guan.
\newblock High dynamic range real-time 3d measurement based on fourier transform profilometry.
\newblock \emph{Optics \& Laser Technology}, 2021.

\bibitem[Long et~al.(2014)Long, Shelhamer, and Darrell]{DBLP:journals/corr/LongSD14}
Jonathan Long, Evan Shelhamer, and Trevor Darrell.
\newblock Fully convolutional networks for semantic segmentation.
\newblock \emph{CoRR}, abs/1411.4038, 2014.

\bibitem[Lu et~al.(2022)Lu, Wu, Zhang, Chen, Li, and Li]{LU2022106873}
Lilian Lu, Zhoujie Wu, Qican Zhang, Chaowen Chen, Yueyang Li, and Fengjiao Li.
\newblock High-efficiency dynamic three-dimensional shape measurement based on misaligned gray-code light.
\newblock \emph{Optics and Lasers in Engineering}, 150:\penalty0 106873, 2022.

\bibitem[Luo et~al.(2016)Luo, Schwing, and Urtasun]{7780983}
Wenjie Luo, Alexander~G. Schwing, and Raquel Urtasun.
\newblock Efficient deep learning for stereo matching.
\newblock In \emph{2016 IEEE Conference on Computer Vision and Pattern Recognition (CVPR)}, pages 5695--5703, 2016.

\bibitem[Mayer et~al.(2016{\natexlab{a}})Mayer, Ilg, Hausser, Fischer, Cremers, Dosovitskiy, and Brox]{Mayer_2016}
Nikolaus Mayer, Eddy Ilg, Philip Hausser, Philipp Fischer, Daniel Cremers, Alexey Dosovitskiy, and Thomas Brox.
\newblock A large dataset to train convolutional networks for disparity, optical flow, and scene flow estimation.
\newblock In \emph{2016 IEEE Conference on Computer Vision and Pattern Recognition (CVPR)}. IEEE, 2016{\natexlab{a}}.

\bibitem[Mayer et~al.(2016{\natexlab{b}})Mayer, Ilg, Hausser, Fischer, Cremers, Dosovitskiy, and Brox]{mayer2016large}
Nikolaus Mayer, Eddy Ilg, Philip Hausser, Philipp Fischer, Daniel Cremers, Alexey Dosovitskiy, and Thomas Brox.
\newblock A large dataset to train convolutional networks for disparity, optical flow, and scene flow estimation.
\newblock In \emph{Proceedings of the IEEE conference on computer vision and pattern recognition}, pages 4040--4048, 2016{\natexlab{b}}.

\bibitem[Ordones et~al.(2021)Ordones, Servin, and Kang]{Ordones:21}
Sotero Ordones, Manuel Servin, and John~S. Kang.
\newblock Moire profilometry through simultaneous dual fringe projection for accurate phase demodulation: a comparative study.
\newblock \emph{Appl. Opt.}, 60\penalty0 (28):\penalty0 8667--8675, 2021.

\bibitem[O’Dowd et~al.(2020)O’Dowd, Wachtor, and Todd]{ODOWD2020106293}
Niall~M. O’Dowd, Adam~J. Wachtor, and Michael~D. Todd.
\newblock A model for describing phase-converted image intensity noise in digital fringe projection techniques.
\newblock \emph{Optics and Lasers in Engineering}, 134:\penalty0 106293, 2020.

\bibitem[Pang et~al.(2018{\natexlab{a}})Pang, Sun, Ren, Yang, and Yan]{pang2018cascade}
Jiahao Pang, Wenxiu Sun, Jimmy~SJ. Ren, Chengxi Yang, and Qiong Yan.
\newblock Cascade residual learning: A two-stage convolutional neural network for stereo matching, 2018{\natexlab{a}}.

\bibitem[Pang et~al.(2018{\natexlab{b}})Pang, Sun, Yang, Ren, Xiao, Zeng, and Lin]{pang2018zoom}
Jiahao Pang, Wenxiu Sun, Chengxi Yang, Jimmy Ren, Ruichao Xiao, Jin Zeng, and Liang Lin.
\newblock Zoom and learn: Generalizing deep stereo matching to novel domains, 2018{\natexlab{b}}.

\bibitem[Ronneberger et~al.(2015)Ronneberger, Fischer, and Brox]{DBLP:journals/corr/RonnebergerFB15}
Olaf Ronneberger, Philipp Fischer, and Thomas Brox.
\newblock U-net: Convolutional networks for biomedical image segmentation.
\newblock \emph{CoRR}, abs/1505.04597, 2015.

\bibitem[Seki and Pollefeys(2016)]{Seki2016PatchBC}
Akihito Seki and Marc Pollefeys.
\newblock Patch based confidence prediction for dense disparity map.
\newblock In \emph{British Machine Vision Conference}, 2016.

\bibitem[Shen et~al.(2021)Shen, Dai, and Rao]{shen2021cfnet}
Zhelun Shen, Yuchao Dai, and Zhibo Rao.
\newblock Cfnet: Cascade and fused cost volume for robust stereo matching, 2021.

\bibitem[Shen et~al.(2022)Shen, Dai, Song, Rao, Zhou, and Zhang]{shen2022pcwnet}
Zhelun Shen, Yuchao Dai, Xibin Song, Zhibo Rao, Dingfu Zhou, and Liangjun Zhang.
\newblock Pcw-net: Pyramid combination and warping cost volume for stereo matching, 2022.

\bibitem[Song et~al.(2016)Song, Hu, Wen, and Yan]{SONG201674}
Kechen Song, Shaopeng Hu, Xin Wen, and Yunhui Yan.
\newblock Fast 3d shape measurement using fourier transform profilometry without phase unwrapping.
\newblock \emph{Optics and Lasers in Engineering}, 84:\penalty0 74--81, 2016.

\bibitem[Song et~al.(2021)Song, Yang, Zhu, Zhou, Ma, Wang, and Shi]{song2021adastereo}
Xiao Song, Guorun Yang, Xinge Zhu, Hui Zhou, Yuexin Ma, Zhe Wang, and Jianping Shi.
\newblock Adastereo: An efficient domain-adaptive stereo matching approach, 2021.

\bibitem[Song et~al.(2019)Song, Wang, Jiang, Liu, and Rao]{DBLP:journals/corr/abs-1902-09314}
Youwei Song, Jiahai Wang, Tao Jiang, Zhiyue Liu, and Yanghui Rao.
\newblock Attentional encoder network for targeted sentiment classification.
\newblock \emph{CoRR}, abs/1902.09314, 2019.

\bibitem[Su et~al.(2019)Su, Gao, Fang, Liu, Wang, Zhang, and Wu]{Su2019TheoreticalAO}
Yong Su, Zeren Gao, Zheng Fang, Yang Liu, Yaru Wang, Qingchuan Zhang, and Shangquan Wu.
\newblock Theoretical analysis on performance of digital speckle pattern: uniqueness, accuracy, precision, and spatial resolution.
\newblock \emph{Optics express}, 27 16:\penalty0 22439--22474, 2019.

\bibitem[Taniai et~al.(2017)Taniai, Matsushita, Sato, and Naemura]{taniai2017continuous}
Tatsunori Taniai, Yasuyuki Matsushita, Yoichi Sato, and Takeshi Naemura.
\newblock Continuous 3d label stereo matching using local expansion moves.
\newblock \emph{IEEE transactions on pattern analysis and machine intelligence}, 40\penalty0 (11):\penalty0 2725--2739, 2017.

\bibitem[Tankovich et~al.(2023)Tankovich, Häne, Zhang, Kowdle, Fanello, and Bouaziz]{tankovich2023hitnet}
Vladimir Tankovich, Christian Häne, Yinda Zhang, Adarsh Kowdle, Sean Fanello, and Sofien Bouaziz.
\newblock Hitnet: Hierarchical iterative tile refinement network for real-time stereo matching, 2023.

\bibitem[Tonioni et~al.(2019)Tonioni, Rahnama, Joy, Stefano, Ajanthan, and Torr]{tonioni2019learning}
Alessio Tonioni, Oscar Rahnama, Thomas Joy, Luigi~Di Stefano, Thalaiyasingam Ajanthan, and Philip H.~S. Torr.
\newblock Learning to adapt for stereo, 2019.

\bibitem[Tosi et~al.(2021)Tosi, Liao, Schmitt, and Geiger]{tosi2021smdnets}
Fabio Tosi, Yiyi Liao, Carolin Schmitt, and Andreas Geiger.
\newblock Smd-nets: Stereo mixture density networks, 2021.

\bibitem[{Van Crombrugge} et~al.(2020){Van Crombrugge}, Penne, and Vanlanduit]{VANCROMBRUGGE2020106305}
Izaak {Van Crombrugge}, Rudi Penne, and Steve Vanlanduit.
\newblock Extrinsic camera calibration for non-overlapping cameras with gray code projection.
\newblock \emph{Optics and Lasers in Engineering}, 134:\penalty0 106305, 2020.

\bibitem[{Van der Jeught} and Dirckx(2016)]{VANDERJEUGHT201618}
Sam {Van der Jeught} and Joris~J.J. Dirckx.
\newblock Real-time structured light profilometry: a review.
\newblock \emph{Optics and Lasers in Engineering}, 87:\penalty0 18--31, 2016.
\newblock Digital optical \& Imaging methods in structural mechanics.

\bibitem[Wan and Kong(2023)]{Wan:23}
MingZhu Wan and Lingbao Kong.
\newblock Single-shot 3d measurement of highly reflective objects with deep learning.
\newblock \emph{Opt. Express}, 31\penalty0 (9):\penalty0 14965--14985, 2023.

\bibitem[Wang et~al.(2021{\natexlab{a}})Wang, Fan, Cai, and Liu]{Wang_2021}
Hengli Wang, Rui Fan, Peide Cai, and Ming Liu.
\newblock Pvstereo: Pyramid voting module for end-to-end self-supervised stereo matching.
\newblock \emph{IEEE Robotics and Automation Letters}, 6\penalty0 (3):\penalty0 4353–4360, 2021{\natexlab{a}}.

\bibitem[Wang et~al.(2021{\natexlab{b}})Wang, Fan, and Liu]{wang2021scvstereo}
Hengli Wang, Rui Fan, and Ming Liu.
\newblock Scv-stereo: Learning stereo matching from a sparse cost volume, 2021{\natexlab{b}}.

\bibitem[Wang et~al.(2020{\natexlab{a}})Wang, Jampani, Sun, Loop, Birchfield, and Kautz]{wang2020improving}
Jialiang Wang, Varun Jampani, Deqing Sun, Charles Loop, Stan Birchfield, and Jan Kautz.
\newblock Improving deep stereo network generalization with geometric priors, 2020{\natexlab{a}}.

\bibitem[Wang et~al.(2021{\natexlab{c}})Wang, Cao, Li, Wan, Li, Xu, and Zhang]{Wang:21}
Lu Wang, Yiping Cao, Chengmeng Li, Yingying Wan, Hongmei Li, Cai Xu, and Hechen Zhang.
\newblock Improved computer-generated moir\'{e} profilometry with flat image calibration.
\newblock \emph{Appl. Opt.}, 60\penalty0 (5):\penalty0 1209--1216, 2021{\natexlab{c}}.

\bibitem[Wang et~al.(2017)Wang, Ri, Tsuda, Koyama, and Tsuzaki]{Wang:17}
Qinghua Wang, Shien Ri, Hiroshi Tsuda, Motomichi Koyama, and Kaneaki Tsuzaki.
\newblock Two-dimensional moir\&\#x00e9; phase analysis for accurate strain distribution measurement and application in crack prediction.
\newblock \emph{Opt. Express}, 25\penalty0 (12):\penalty0 13465--13480, 2017.

\bibitem[Wang et~al.(2020{\natexlab{b}})Wang, Okumura, Ri, Xia, Tsuda, and Ogihara]{Wang:20}
Qinghua Wang, Shigesato Okumura, Shien Ri, Peng Xia, Hiroshi Tsuda, and Shinji Ogihara.
\newblock Second-order moir\&\#x00e9; method for accurate deformation measurement with a large field of view.
\newblock \emph{Opt. Express}, 28\penalty0 (5):\penalty0 7498--7514, 2020{\natexlab{b}}.

\bibitem[Wang et~al.(2024)Wang, Xu, Jia, and Yang]{Wang_2024_CVPR}
Xianqi Wang, Gangwei Xu, Hao Jia, and Xin Yang.
\newblock Selective-stereo: Adaptive frequency information selection for stereo matching.
\newblock In \emph{Proceedings of the IEEE/CVF Conference on Computer Vision and Pattern Recognition (CVPR)}, pages 19701--19710, 2024.

\bibitem[Wu et~al.(2022{\natexlab{a}})Wu, Cao, An, Li, Li, Xu, and Yang]{WU2022107004}
Haitao Wu, Yiping Cao, Haihua An, Yang Li, Hongmei Li, Cai Xu, and Na Yang.
\newblock A novel phase-shifting profilometry to realize temporal phase unwrapping simultaneously with the least fringe patterns.
\newblock \emph{Optics and Lasers in Engineering}, 153:\penalty0 107004, 2022{\natexlab{a}}.

\bibitem[Wu et~al.(2022{\natexlab{b}})Wu, Cao, An, Xu, Li, and Li]{WU2022107955}
Haitao Wu, Yiping Cao, Haihua An, Cai Xu, Hongmei Li, and Yang Li.
\newblock A general phase ambiguity suppression algorithm combining geometric constraints and temporal phase unwrapping.
\newblock \emph{Optics \& Laser Technology}, 150:\penalty0 107955, 2022{\natexlab{b}}.

\bibitem[Wu et~al.(2019)Wu, Wu, Zhang, Wang, and Ju]{wu2019semantic}
Zhenyao Wu, Xinyi Wu, Xiaoping Zhang, Song Wang, and Lili Ju.
\newblock Semantic stereo matching with pyramid cost volumes.
\newblock In \emph{Proceedings of the IEEE/CVF international conference on computer vision}, pages 7484--7493, 2019.

\bibitem[Wu et~al.(2020)Wu, Guo, Li, Liu, and Zhang]{article03}
Zhoujie Wu, Wenbo Guo, Yueyang Li, Yihang Liu, and Qican Zhang.
\newblock High-speed and high-efficiency three-dimensional shape measurement based on gray-coded light.
\newblock \emph{Photonics Research}, 8, 2020.

\bibitem[Xiao and Li(2017)]{XIAO201719}
Yi Xiao and Youfu Li.
\newblock High-quality binary fringe generation via joint optimization on intensity and phase.
\newblock \emph{Optics and Lasers in Engineering}, 97:\penalty0 19--26, 2017.

\bibitem[Xu et~al.(2021)Xu, Xu, Yang, Jia, and Guo]{xu2021bilateral}
Bin Xu, Yuhua Xu, Xiaoli Yang, Wei Jia, and Yulan Guo.
\newblock Bilateral grid learning for stereo matching networks, 2021.

\bibitem[Xu et~al.(2022)Xu, Cheng, Guo, and Yang]{xu2022attention}
Gangwei Xu, Junda Cheng, Peng Guo, and Xin Yang.
\newblock Attention concatenation volume for accurate and efficient stereo matching, 2022.

\bibitem[Xu et~al.(2023)Xu, Wang, Ding, and Yang]{xu2023iterative}
Gangwei Xu, Xianqi Wang, Xiaohuan Ding, and Xin Yang.
\newblock Iterative geometry encoding volume for stereo matching, 2023.

\bibitem[Xu and Zhang(2020)]{xu2020aanet}
Haofei Xu and Juyong Zhang.
\newblock Aanet: Adaptive aggregation network for efficient stereo matching, 2020.

\bibitem[Yu et~al.(2022)Yu, Gong, Wu, Sun, Zhao, Wu, and Yu]{YU2022128236}
Shuang Yu, Ting Gong, Haibin Wu, Xiaoming Sun, Yanqiao Zhao, Shuang Wu, and Xiaoyang Yu.
\newblock 3d shape measurement based on the unequal-period combination of shifting gray code and dual-frequency phase-shifting fringes.
\newblock \emph{Optics Communications}, 516:\penalty0 128236, 2022.

\bibitem[Zhang et~al.(2019)Zhang, Prisacariu, Yang, and Torr]{zhang2019ganet}
Feihu Zhang, Victor Prisacariu, Ruigang Yang, and Philip H.~S. Torr.
\newblock Ga-net: Guided aggregation net for end-to-end stereo matching, 2019.

\bibitem[Zhang et~al.(2020)Zhang, Qi, Yang, Prisacariu, Wah, and Torr]{zhang2020domain}
Feihu Zhang, Xiaojuan Qi, Ruigang Yang, Victor Prisacariu, Benjamin Wah, and Philip Torr.
\newblock Domain-invariant stereo matching networks.
\newblock In \emph{Computer Vision--ECCV 2020: 16th European Conference, Glasgow, UK, August 23--28, 2020, Proceedings, Part II 16}, pages 420--439. Springer, 2020.

\bibitem[Zhang et~al.(2024)Zhang, Li, Huang, Yu, Gu, Zheng, and Bai]{zhang2024robust}
Jiawei Zhang, Jiahe Li, Lei Huang, Xiaohan Yu, Lin Gu, Jin Zheng, and Xiao Bai.
\newblock Robust synthetic-to-real transfer for stereo matching.
\newblock In \emph{Proceedings of the IEEE/CVF Conference on Computer Vision and Pattern Recognition}, pages 20247--20257, 2024.

\bibitem[Zhao et~al.(2022)Zhao, Zhou, Zhang, Zhao, Yang, and Ouyang]{Zhao2022EAIStereoEA}
Haoliang Zhao, Huizhou Zhou, Yongjun Zhang, Yong Zhao, Yitong Yang, and Ting Ouyang.
\newblock Eai-stereo: Error aware iterative network for stereo matching.
\newblock In \emph{Asian Conference on Computer Vision}, 2022.

\bibitem[Zhao et~al.(2023)Zhao, Zhou, Zhang, Chen, Yang, and Zhao]{10203487}
Haoliang Zhao, Huizhou Zhou, Yongjun Zhang, Jie Chen, Yitong Yang, and Yong Zhao.
\newblock High-frequency stereo matching network.
\newblock In \emph{2023 IEEE/CVF Conference on Computer Vision and Pattern Recognition (CVPR)}, pages 1327--1336, 2023.

\bibitem[Zheng et~al.(2017)Zheng, Da, Kemao, and Seah]{Zheng:17}
Dongliang Zheng, Feipeng Da, Qian Kemao, and Hock~Soon Seah.
\newblock Phase-shifting profilometry combined with gray-code patterns projection: unwrapping error removal by an adaptive median filter.
\newblock \emph{Opt. Express}, 25\penalty0 (5):\penalty0 4700--4713, 2017.

\bibitem[Zhou et~al.(2018)Zhou, Zhu, and Jing]{article01}
Pei Zhou, Jiangping Zhu, and Hailong Jing.
\newblock Optical 3-d surface reconstruction with color binary speckle pattern encoding.
\newblock \emph{Optics Express}, 26:\penalty0 3452, 2018.

\bibitem[Zhu et~al.(2023)Zhu, Li, Zhou, and You]{ZHU2023107542}
Zhenmin Zhu, Minchao Li, Fuqiang Zhou, and Duoduo You.
\newblock Stable 3d measurement method for high dynamic range surfaces based on fringe projection profilometry.
\newblock \emph{Optics and Lasers in Engineering}, 166:\penalty0 107542, 2023.

\bibitem[Zuo et~al.(2021)Zuo, Yin, Hu, Feng, HUANG, Qian, and Chen]{article02}
Chao Zuo, Wei Yin, Yan Hu, Shijie Feng, Lei HUANG, Kemao Qian, and Qian Chen.
\newblock Single-shot 3d shape measurement using an end-to-end stereo matching network for speckle projection profilometry.
\newblock \emph{Optics Express}, 29, 2021.

\end{thebibliography}
}

\end{document}